\documentstyle[aps]{revtex}
\begin{document}
\title{Influence Action and Decoherence of Hydrodynamic Modes }
\author{E. A. Calzetta \thanks{%
Electronic address: {\tt calzetta@df.uba.ar}} $^1$ and B. L. Hu \thanks{%
Electronic address: {\tt hub@physics.umd.edu}} $^2$}
\address{$^1$ Department of Physics and IAFE, University of Buenos Aires, Argentina}
\address{$^2$ Department of Physics, University of Maryland,\\
College Park, Maryland 20742-4111}
\date{{\it umdpp 99-025, submitted to Physical Review D Sept. 27, 1998}}
\maketitle

\begin{abstract}
We derive an influence action for the heat diffusion equation and from its
spectral dependence show that long wavelength hydrodynamic modes are more
readily decohered. The result is independent of the details of the
microscopic dynamics, and follows from general principles alone.
\end{abstract}

%\date{\today}

In the discussion of classical equations from quantum dynamics, Gell-Mann
and Hartle \cite{GelHar2} pointed out that for a large and possibly complex
system the variables that will become classical `habitually' are the local
densities integrated over small volumes. To show that some variables become
classical involves showing that they are readily decoherent and that they
obey deterministic evolution equations. Hydrodynamic variables such as
energy, momentum and number are of such characters because they are
conserved quantities. Interesting work has been done in recent years by
Hartle, Halliwell and co-workers \cite{DecHyd} in applying the decoherent
history approach \cite{conhis} to systems consisting of large numbers of
particles with histories consisting of projections onto local densities. The
task is to show that these histories decohere and that their probabilities
are peaked about hydrodynamic equations.

For a large system consisting of many particles of equal importance, such as
the Boltzmann gas, the traditional environment-induced decoherence approach 
\cite{envdec} assuming a distinguished system such as the quantum Brownian
motion (QBM) \cite{if} becomes inoperative. Among a number of alternatives,
the present authors suggested using the correlation hierarchy to represent
the complete system and aiming at the decoherence of histories defined in
terms of (n-th order) correlation functions \cite{dch,charis}. To
investigate the decoherence of hydrodynamic variables in this vein, the task
would be to study the hydrodynamic limit of the correlation hierarchy and
show that such limit indeed possesses special decoherent characteristics to
warrant emergent classicality. The first step has been a long standing issue
in the foundation of statistical mechanics (see e.g., \cite{hydro}); the
second step in correlation decoherence is a new challenge.

We are not prepared to address the above problem as yet, but wish to offer
an observation in this brief note on why hydrodynamic modes are most readily
decohered from the viewpoint of the closed-time-path (CTP) \cite{ctp,CH87}
or influence functional \cite{if} formalism. In our previous work we (as
well as others) have shown the intimate relation between the CTP and the
influence functional (IF) \cite{CH94}, and between the IF and the
decoherence functional (DF) \cite{dch,envdechis}. Because of the way it is
set up, i.e., in terms of open systems with a clear system- environment
split, the influence functional is usually incorrectly viewed as unsuitable
for effectively closed (or effectively open) systems. By an effectively open
system we mean systems where the environment is not clearly identifiable (as
is in the QBM) but effectively exists and exerts an influence on the system
just like an open system. The influence functional is equally applicable to
these cases. \footnote{%
We can see this via its relation with the CTP effective action, which
traditionally has been used for the derivation of evolutionary equations for
particle processes, with no obvious introduction of an environment. But, as
we pointed out earlier \cite{CH97}, the presence of a quantum field
constitutes a de facto environment, and the equations are effective in the
sense that radiative corrections have been included.} There are many ways
how an effectively open system can be defined, using a discrepancy between
fast-slow variables, long-short wavelength excitations, or high-low energy
scale, etc \cite{cgea}. The thermodynamics-hydrodynamics regime (of a closed
system like the Boltzmann gas) in question here belongs to this latter
category \cite{molhyd}. Indeed, for our problem, there is an environment
(from, say, the presence of other constituents or the short wavelength
sector of the system) which defines the temperature but in the linear
response regime no particular coupling or any of its microscopic features
need be specified or will manifest. This is what makes our derivation
possible, a point which will be made clear in our result.

To show the decoherent properties we need to first identify from the
(coarse-grained, large scale) hydrodynamic equations the noise arising from
and reflecting its microscopic constituents. We know that noise (in the
environment, or effective environment) is instrumental to the decoherence of
the system. \footnote{%
Physically, nearly conserved quantities, and hydrodynamic modes in
particular, are usually only weakly coupled to the rest of the universe --
that is why they are nearly conserved. There is a delicate balance between
having just enough noise to decoherence but not in excess so as to corrupt
the deterministic (classical) path. See \cite{GelHar2}} We do this with the
help of the fluctuation-dissipation relation (FDR) \cite{fdr}, which relates
the dissipation in the long wavelength modes with their fluctuations. The
FDR gives the auto-correlation of the noise. From this noise one can
construct a Langevin equation governing the dissipation of thermodynamic
perturbations (in our example the temperature) which describes the approach
to thermal equilibrium. Added to this -- that the influence action should
give the Langevin dynamics upon variation-, we demand also that it should
produce the finite temperature free energy density when time is continued
into the imaginary domain (with periodicity given by the inverse
temperature). These two conditions are enough to determine its structure.
Finally, reinterpreting the influence action as essentially the logarithm of
the decoherence functional, we are in a position to decide which modes of
the temperature field are most readily decohered. As could be expected,
these are the long wavelength, hydrodynamic modes.

Our general argument complements the explicit models presented in Ref. \cite
{DecHyd}. Rather than explicitly constructing a nearly conserved quantity
and showing its decoherence, we emphasize the generic properties leading to
the classicality of the hydrodynamic modes. As we shall see, decoherence of
the hydrodynamic modes is, at least for systems close to equilibrium,
essentially a consequence of the Second Law and the Einstein relation, which
is a fluctuation-dissipation relation for linear responses. \footnote{%
We do not claim that the consistent histories approach in any way forces us
to consider histories defined in terms of hydrodynamical variables;
certainly other consistent sets are also possible, maybe some even
contradictory to the hydro histories \cite{kent}. We only wish to point out
that the good properties of histories defined in terms of nearly conserved
quantities vis a vis decoherence are not an accident, but rather follow from
the phenomenology of macroscopic behavior, as encoded in the Laws of
Thermodynamics. As such this kind of histories constitutes an {\it %
interesting} classical domain \cite{dch}; whether we should use this domain
to describe our actual experience is a matter of physics, not formalism.}

To focus on the basic issues, we shall look into the simplest case, that of
a nonrelativistic, heat conducting medium with no matter motion. We shall
consider a linear perturbation $\delta T$ in the temperature field from some
background, uniform temperature $T_0$. As is well known, the perturbation
will regress to equilibrium, and this process is described by a Langevin
equation. We assume that the only relevant thermodynamic variable is the
local temperature $T=T_0+\delta T\left(\vec x,t\right) $, everything else
having been coarse grained. The local temperature is associated to the
variation in energy density $u$ in the usual way

\begin{equation}
du\left(\vec x\right) =c\left( T\right) dT\left(\vec x\right)
\label{specificheat}
\end{equation}
where $c=c(T)$ is the specific heat. The relevant conservation law is the
First Law of Thermodynamics (we use the convention that positive heat means
flow into the body, see \cite{Fermi}), $u_{,t}=-\vec \nabla \vec q$, where $%
\vec q$ is the heat flux. Near equilibrium the heat flux is given by
Fourier's Law (which is enough in the non relativistic theory) $\vec q=-D%
\vec \nabla T$, where $D$ is the heat diffusion constant \cite{degroot}.
Since we only consider the linear response, we may work with a single
Fourier mode $\delta T_k\left( t\right) $. The macroscopic evolution
equation is

\begin{equation}
c_0\frac \partial {\partial t}(\delta T_k)+D_0k^2(\delta T_k)=0
\label{detereq}
\end{equation}
where $c_0=c\left( T_0\right) $, and similarly $D_0=D\left( T_0\right) .$
Roughly speaking, we can use $k^2$ to gauge conservation, with better
conservation for smaller wavenumbers.

Near equilibrium, the local temperature will undergo spontaneous
fluctuations, with a Gaussian probability density $\rho \sim \exp -\beta
_0\Delta F$, where $\beta _0=1/T_0,$ and $\Delta F=\Delta U-T_0\Delta S$ is
the free energy, $\Delta U,$ $\Delta S$ being the total energy and entropy
change associated to the fluctuation. It is crucial for our argument that
the free energy depends on the global temperature $T_0,$ rather than on the
local temperature $T=T_0+\delta T\left( \vec x,t\right) .$ For a system in
canonical equilibrium at temperature $T_0$, the probability of a microscopic
configuration adding up to a local temperature fluctuation $\delta T\left( 
\vec x,t\right) $ is $\exp -\beta _0U$, while the number of such
configurations is $\exp S$ , so, after normalizing by the factor $\exp \beta
_0F_0$, where $F_0=U_0-T_0S_0$ is the equilibrium free energy, we obtain the
total probability as given. We obtain the same result if we apply Einstein's
formula to the closed system made up of our system and the heat reservoir.

We wish to compute the change in the free energy as a result of a
temperature fluctuation. Mathematically it is important to keep in mind the
distinction between an extensive quantity and its density, on one hand, and
the variational derivative of the former and the ordinary derivatives of the
latter, on the other. For example, define $U=\int d^dx\;u\left(\vec x\right)$
then Eq. (\ref{specificheat}) can be read as

\begin{equation}
\frac{\delta U}{\delta T\left(\vec x\right) }=\left. \frac{\partial u}{%
\partial T}\right| _{T=T\left(\vec x\right) }=c\left[ T\left(\vec x\right)
\right]
\end{equation}
where $\delta $ denotes a variational derivative, $\partial $ a partial one.

The variations of the different quantities are constrained by thermodynamic
laws; for example, we have

\[
ds\left(\vec x\right) =\frac{du\left(\vec x\right) }{T\left(\vec x\right) } 
\]
where $s$ is the entropy density. Combining these we obtain the identity
relating the first variations of energy and entropy

\begin{equation}
\left. \frac{\delta S}{\delta T\left(\vec x\right) }\right| _{T=T_0}=\frac 1{%
T_0}\left. \frac{\delta U}{\delta T\left(\vec x\right) }\right| _{T=T_0}
\label{primovar}
\end{equation}
which in particular implies that the first variation of the free energy
vanishes in equilibrium, as expected.

For the second variation we get

\begin{equation}
\frac{\delta ^2F}{\delta T\left(\vec x\right) \delta T\left(\vec y\right) }=%
\frac{\delta ^2U}{\delta T\left(\vec x\right) \delta T\left(\vec y\right) }%
-T_0\frac{\delta ^2S}{\delta T\left(\vec x\right) \delta T\left(\vec y%
\right) }  \label{secondvar}
\end{equation}
But

\begin{equation}
\frac{\delta ^2S}{\delta T\left(\vec x\right) \delta T\left(\vec y\right) }=%
\frac \delta {\delta T\left(\vec x\right) }\left[ \frac{\delta S}{\delta
T\left(\vec y\right) }\right] =\frac \delta {\delta T\left(\vec x\right) }%
\left[ \frac 1{T\left(\vec y\right) }\frac{\delta U}{\delta T\left(\vec y%
\right) }\right]  \label{aux}
\end{equation}
so the second variation of the free energy is 
\begin{equation}
\frac{\delta ^2F}{\delta T\left(\vec x\right) \delta T\left(\vec y\right) }=%
\frac 1{T_0}\frac{\delta U}{\delta T\left(\vec y\right) }\delta \left(\vec x%
- \vec y\right) =\frac 1{T_0}\left. \frac{\partial u}{\partial T}\right|
_{T=T_0}\delta \left(\vec x- \vec y\right) =\frac{c_0}{T_0}\delta \left(\vec 
x- \vec y\right)  \label{secondvar2}
\end{equation}
and the change in free energy due to this temperature fluctuation is finally

\begin{equation}
\Delta F=\int d^dx\;\left\{ \frac{c_0}{2T_0}(\delta T)^2+...\right\}
\label{fenchange}
\end{equation}

It follows that in equilibrium each mode $\delta T_k$ has a gaussian
probability distribution, with $<(\delta T_k)^2>=T_0^2/c_0.$ As a check, the
rms value of the energy fluctuation is

\begin{equation}
\left\langle \left( \Delta U\right) ^2\right\rangle =\int
d^dxd^dy\;\left\langle \Delta u\left(\vec x\right) \Delta u\left(\vec y%
\right) \right\rangle =C_0T_0^2
\end{equation}
where $C=Vc$ is the heat capacity, as it should \cite{callen}.

As reasoned above, the fluctuation-dissipation relation requires that
Equation (\ref{detereq}) be supplemented by a noise source on the right hand
side, to support these fluctuations. The microscopic equation of motion is
then

\begin{equation}
\frac \partial {\partial t}(\delta T_k)+\frac{D_0k^2}{c_0}(\delta T_k)=\xi _k
\label{langevin}
\end{equation}
where the $\xi _k$'s are white noise with correlation $2\Gamma _k$:

\begin{equation}
<\xi _k\left( t_1\right) \xi _k\left( t_2\right) >=2\Gamma _k\delta \left(
t_1-t_2\right) ;\quad \Gamma _k=\frac{D_0k^2T_0^2}{c_0^2}  \label{fdt}
\end{equation}

We wish to write down a CTP effective action or influence action ${\cal A}%
_{IF}$ (related to the influence functional ${\cal F}$ by ${\cal F}\equiv
e^{i{\cal A_{IF}}}$) designed to reproduce Eq. (\ref{langevin}). ${\cal A}%
_{IF}$ is a functional of two (rather than one, as in an ordinary action
functional) thermal histories $(\delta T)^1$ and $(\delta T)^2$; this is
such as to enable us to formulate a non time- reversal- invariant causal
theory within a variational principle \cite{CH87}. The symmetry property $%
{\cal A}_{IF}\left[ (\delta T)^1,(\delta T)^2\right] =-{\cal A}_{IF}\left[
(\delta T)^2,(\delta T)^1\right] ^{*}$ implies that the real (imaginary)
part is odd (even) in $(\delta T)^1-(\delta T)^2.$ ${\cal A}_{IF}$ must have
the structure 
\begin{equation}
{\cal A}_{IF}\left[ (\delta T)^1,(\delta T)^2\right] =\frac 12\int d^dk
dtdt^{\prime }\;\left[ \delta T_k\right] (t){\mu_k }(t,t^{\prime })\left\{
\delta T_k\right\} (t^{\prime })+\frac i2\int d^dk dtdt^{\prime }\;\left[
\delta T_k\right] (t)\nu_k(t,t^{\prime })\left[ \delta T_k\right] (t^{\prime
})  \label{ctpea}
\end{equation}
where $\mu_k$ and $\nu_k$ are the dissipation and noise kernels,
respectively, and $[],\{\}$ around a quantity denote taking the difference
and sum of the CTP ($1,2$) components. 
%(In the above an integration $\int d^dk$ is understood; 
%we shall omit it and drop the subscripts $k$ on $\delta T, 
%\xi, \mu,\nu$ to avoid notational overload.)
This functional leads to the equations of motion (see \cite{if})

\begin{equation}
\int dt^{\prime }\;{\mu }_k(t,t^{\prime })\delta T_k(t^{\prime })=\xi
_k\left( t\right) ;\quad \left\langle \xi _k\left( t\right) \xi _k\left(
t^{\prime }\right) \right\rangle =\nu _k(t,t^{\prime })  \label{eqsmot}
\end{equation}

Comparing Equations (\ref{eqsmot}) to (\ref{langevin}) and (\ref{fdt}), we
conclude the influence functional should have the form

\begin{equation}
{\cal A}_{IF}=\frac 12\int d^dk dt\;\left( \left[ \delta T_k\right]
A_k\right) \left\{ \frac \partial {\partial t}+\frac{D_0k^2}{c_0}\right\}
\left\{ \delta T_k\right\} +i\int d^dk dt\;\frac{D_0k^2T_0^2}{c_0^2}\left(
A_k\left[ \delta T_k\right] \right) ^2  \label{ctp}
\end{equation}
In principle, $A_k$ could be any nonsingular operator, but it is simplest to
assume it is local and time independent. We determine $A_k$ by requesting
that, for time independent configurations, the real part of the CTP
effective action, when rotated into imaginary time and integrated from $0$
to $-i\beta _0,$ should reduce to $i\beta _0\left( F\left[ (\delta
T)^1\right] -F\left[ (\delta T)^2\right] \right) $, where $F$ is the free
energy Eq. (\ref{fenchange}). Thus

\begin{equation}
A_k=\frac{c_0^2}{D_0k^2T_0} = \frac {T_0}{\Gamma_k}  \label{constant}
\end{equation}
Finally using the relation between the influence or CTP functional ${\cal F}$
and the decoherence functional ${\cal D}$ \cite{dch,CH94}, we get for the DF:

\begin{equation}
\left| {\cal D}\left[ (\delta T)^1,(\delta T)^2\right] \right| ^2\sim \exp
\left\{ -\int d^dkdt\;\frac{2c_0^2}{D_0k^2}\left[ \delta T_k\right]
^2\right\}   \label{deco}
\end{equation}

We see that indeed the long wavelength modes are the most efficiently
decohered, in agreement with the espoused ideas that maximal decoherence
would pertain to the conserved quantities \cite{GelHar2}. In particular, we
recover the awaited result that conserved quantities are exactly decohered.

It is of interest that we have been able to write down the influence (or CTP
effective) action without seemingly making any assumption concerning the
structure and dynamics of the environment. In reality, there is of course an
environment which provides the finite temperature background. However, here
we study only weak perturbations in the linear response regime, and for weak
linear couplings the transport functions are independent of the microscopic
details of the environment. This subtle yet important observation was made
in a footnote of the paper by Feynman and Vernon which captures the
microscopic theoretical basis for linear response theory.

In conclusion, our result suggests that decoherence of the hydrodynamic
modes is, at least for systems close to equilibrium, essentially a
consequence of the Second Law, the Fourier Law, and the
Fluctuation-Dissipation Theorem.\\

{\bf Acknowledgements} This work began when both authors were at the Santa
Fe Workshop on nonequilibrium phase transitions and concluded at the Third
Peyresq Meeting on Cosmology. We thank Drs. Emil Mottola and Edgard Gunzig
for providing us with a pleasant work environment and to Dr. J. J. Halliwell
for multiple discussions. EC is partially supported by CONICET, UBA and
Fundaci\'on Antorchas (Argentina). BLH is supported in part by the National
Science Foundation (NSF) under grant PHY 98-00967. This collaboration is
partially supported by CONICET and NSF grant INT 95-09847 under the
Scientific and Technological Exchange Program between Argentina and U.S.A.

%\newpage

\end{document}